\documentclass{ws-procs975x65}
\usepackage{verbatim}
\def\ba{\begin{eqnarray}}
\def\ea{\end{eqnarray}}

\begin{document}

\title{Is there weak birefringence of light in vacuo?}
\author{Nadine Stritzelberger}
\address{Department of Applied Mathematics, University of Waterloo,
200 University Ave W,\\ Waterloo, ON, Canada N2L 3G1}
\address{Perimeter Institute for Theoretical Physics, 31 Caroline Street North,\\
Waterloo, Ontario, Canada N2L 2Y5}
\address{Institute for Quantum Computing, University of Waterloo,
200 University Ave W,\\ Waterloo, ON, Canada N2L 3G1\\
E-mail: nadine.stritzelberger@cantab.net}

\begin{abstract}
There are numerous reasons to study modifications of general relativity and the Standard Model of particle physics, ranging from modelling inflation to exploring galaxy rotation curves and the nature of dark matter. 
Here we study the most general linear theory of electromagnetism, which admits vacuum birefringence, and derive weak gravitational field equations for the underlying area metric spacetime geometry. We discuss the weak gravitational field sourced by a point mass in an area metric spacetime and find first order corrections to the linearized Schwarzschild metric of general relativity.
\end{abstract}


\bodymatter

\section{Introduction}

Could there be weak birefringence of light in vacuo? How can we predict whether and where such splitting of light rays in vacuo occurs?  Standard Maxwell electrodynamics, for which the electromagnetic field couples to a metric spacetime geometry, {\it a priori} excludes the possibility of vacuum birefringence. The spacetime geometry must be refined, from a metric to a tensor field of the fourth rank, in order to obtain a linear theory of electrodynamics allowing for vacuum birefringence to occur. The such obtained matter theory is known as the theory of {\it general linear electrodynamics}\cite{HO,Rubilar1,Rubilar2} and the fourth rank spacetime tensor field as an {\it area metric}.
Because the matter field equations
of general linear electrodynamics depend on the area metric background, finding any concrete solution for the electromagnetic field requires to simultaneously solve some yet-to-be-determined field
equations for the area metric, which thus take the role played by Einstein's field equations for a metric background. 
In particular, also the splitting of light rays
depends crucially on such gravitational field equations underlying general linear electrodynamics. We can thus rephrase the question of whether and where birefringence of light in vacuo occurs as follows: What are the gravitational field equations underlying general linear electrodynamics, and what are the solutions to these field equations? 

\noindent This question can be answered by means of the recently discovered procedure of gravitational closure\cite{DSSW,GSWW}, which employs the causal structure of given matter field equations in order to provide causally compatible canonical dynamics\cite{HKT,K} for the underlying geometry. 
For instance, starting from Maxwell electrodynamics, one obtains the familiar Einstein equations in this way. Exactly along the same lines, when starting instead from general linear electrodynamics, one obtains the gravitational field equations for an area metric geometry.
Based on Ref.~[\refcite{SSSW}], this paper reviews the derivation of the {\it weak} gravitational field equations underlying general linear electrodynamics, followed by a discussion of a concrete solution to these equations, namely the weak gravitational field sourced by a point mass in an area metric spacetime, revealing first order corrections to the linearized Schwarzschild metric.

\section{General linear electrodynamics}

We start our discussion with a brief review of the theory of general linear electrodynamics\cite{HO,Rubilar1,Rubilar2,KM,SWW,RRS}. The action of this matter theory is
\ba
S_{matter}[A,G)] &=& -\frac{1}{8}\int_M d^4 x \, \omega_G F_{ab} F_{cd} G^{abcd}\,, \label{Matter_action}
\ea
where $M$ is an orientable four-dimensional spacetime manifold, $F=\text{d}A$ denotes the electromagnetic field strength tensor, the fourth-rank contravariant tensor field $G$ denotes the area metric spacetime geometry, and the scalar density $\omega_G=4!(\epsilon_{abcd}G^{abcd})^{-1}$ is constructed by virtue of the canonical top form density $\epsilon$. 
In a spacetime equipped with a tensorial spacetime geometry, general linear electrodynamics is the most general theory of electrodynamics for which the classical linear superposition principle still holds.
By virtue of the appearance of the fourth-rank tensor field
in the above matter action, the algebraic symmetries
\ba 
G^{abcd}=G^{cdab} \quad \text{and}\quad
G^{abcd}=-G^{bacd}
\ea 
may be assumed without loss of generality.
Consequently, an area metric spacetime geometry features $21$ independent degrees of freedom at each spacetime point, as opposed to only $10$ independent degrees of freedom for a metric spacetime geometry. It is ultimately due to these additional degrees of freedom that general linear electrodynamics allows not only to describe all of Maxwell electrodynamics, but also various new phenomena such as vacuum birefringence. 

\noindent The causal structure of the equations of motion of general linear electrodynamics is prescribed by the principal tensor field $P_G$, which was shown in Ref. [\refcite{Rubilar1,Rubilar2}] to be a totally symmetric tensor field density of the fourth rank, determined entirely in terms of the spacetime geometry,
\ba 
P_G^{\,abcd} &=& -\frac{1}{4!}\omega_G^2\, \epsilon_{mnpq}\epsilon_{rstu}G^{mnr(a}G^{b|ps|c}G^{d)qtu} \label{principal_polynomial} \,.
\ea
The principal tensor field is the reason why generic area metric spacetime geometries feature vacuum birefringence: The null cone of general linear electrodynamics, that is the set of covectors annihilating the principal tensor field, 
is a quartic surface. 
However, if the splitting of light rays in vacuo was a large effect, it would most likely have been observed by now. 
Therefore, instead of studying generic area metric spacetimes, this paper focuses on spacetime geometries which are small area metric perturbations $H$ around metric Minkowski spacetime $\eta$,
\ba 
G^{abcd} = \eta^{ac}\eta^{bd}-\eta^{ad}\eta^{bc}-\epsilon^{abcd}+H^{abcd}\,. 
\label{area_metric_perturbations}
\ea
Once we know the linearized gravitational field equations underlying general linear electrodynamics, we can derive the area metric perturbations $H$ for given small matter distributions.
In the following two sections, we therefore discuss the derivation of these linearized gravitational field equations by means of gravitational closure.

\section{Canonical geometry}

The procedure of gravitational closure enables the derivation of causally compatible gravitational field equations for the spacetime geometry underlying any predictive and quantizable matter field theory\cite{HKT,K,GSWW,DSSW}.
The first step in this gravitational closure procedure is to determine the canonical geometry for which one ultimately wishes to obtain canonical dynamical field equations. 
The canonical geometry of an area metric spacetime geometry can be determined as follows. Let $\Sigma$ be a three-dimensional manifold and let $X_t:\Sigma \hookrightarrow M$ be a one-real-parameter family of embedding maps, which specifies a foliation of the spacetime manifold into initial data surfaces. For each point $\sigma\in\Sigma$ and each embedding parameter $t\in\mathbb{R}$, one can construct an orthonormal frame $\{\epsilon^0(t,\sigma), \epsilon^\alpha(t,\sigma) \}$,
where $\alpha\in \{1,2,3\}$, which allows to project the area metric spacetime geometry onto the initial data surfaces,
\ba
(g_1)^{\alpha\beta}[X_t] &:=& -G(\epsilon^0, \epsilon^{\alpha}, \epsilon^0, \epsilon^{\beta}) 
~,\\
(g_2)_{\alpha\beta}[X_t] &:=& \frac{1}{4} \frac{\epsilon_{\alpha\mu\nu}}{\sqrt{\det (g_1)} } \frac{\epsilon_{\beta\sigma\tau}}{\sqrt{\det (g_1)}} G(\epsilon^\mu, \epsilon^{\nu}, \epsilon^\sigma, \epsilon^{\tau}) 
~,
\\
(g_3)^{\alpha}{}_{\beta}[X_t] &:=& \frac{1}{2} \frac{\epsilon_{\beta\mu\nu}}{\det (g_1)} G(\epsilon^0, \epsilon^{\alpha}, \epsilon^\mu, \epsilon^{\nu}) - \delta^{\alpha}_{\beta}
~.
\ea
resulting in three one-real-parameter families of induced tenor fields. By construction, the induced tensor fields $g_1$ and $g_2$ are symmetric,
\ba
(g_1)^{[\alpha\beta]}=0 \quad \text{and}\quad 
(g_2)_{[\alpha\beta]}=0 ~.
\label{symmetry_conditions}
\ea 
Moreover, the orthonormality conditions satisfied by the  four frame fields imply
\ba
(g_1)^{\sigma[\alpha}(g_3)^{\beta]}{}_{\sigma}=0 
\quad \text{and} 
\quad 
(g_3)^{\alpha}{}_{\alpha}=0 \,. \label{frame_conditions}
\ea
Therefore, only $17$ of the components of the induced tensor fields are independent degrees of freedom. In order to obtain the canonical geometry from this induced geometry, we need to parametrize the three induced tensor fields $\{g_1,g_2,g_3\}$ in terms of unconstrained canonical configuration variables $\{\varphi^1,\dots,\varphi^{17}\}$. 
Note that since the four frame conditions (\ref{frame_conditions}) are non-linear, any parametrization $\{g_1(\varphi),g_2(\varphi),g_3(\varphi)\}$ respecting the symmetry conditions (\ref{symmetry_conditions}) and frame conditions (\ref{frame_conditions}) is non-linear in the canonical configuration variables. 
One possible parametrization---particularly suited for the perturbative approach pursued in this paper---is the following.
Consider a constant intertwiner $\mathcal{I}^{\alpha\beta}{}_{A}$, distributing the unconstrained configuration variables $\{\varphi^1\,,...\,,\varphi^{17}\}$ over three second rank tensor fields,
\ba
(\varphi_1)^{\alpha\beta} := \sum_{A=1}^{6}\mathcal{I}^{\,\alpha\beta}{}_{A} \varphi^A ,~~~
(\varphi_2)^{\alpha\beta} :=
\sum_{A=7}^{12} \mathcal{I}^{\,\alpha\beta}{}_{A} \varphi^A 
,~~~
(\varphi_3)^{\alpha\beta} := \sum_{A=13}^{17} \mathcal{I}^{\,\alpha\beta}{}_{A} \varphi^A . ~~~~
\ea
Now let $\gamma$ denote a flat three-dimensional Euclidean metric on $\Sigma$ and let the intertwiner $\mathcal{I}^{\,\alpha\beta}{}_{A}$ and its inverse $\mathcal{I}^{A}{}_{\alpha\beta}$ satisfy the completeness relations
\ba
\mathcal{I}^{\alpha\beta}{}_{A}\mathcal{I}^{B}{}_{\alpha\beta} 
= \delta^{B}_{A}
~&,&~~~
\sum_{A=1}^{6} \mathcal{I}^{\alpha\beta}{}_{A} \mathcal{I}^{A}{}_{\mu\nu} 
= \delta^{(\alpha}_{\mu} \delta^{\beta)}_{\nu} 
,\nonumber \\
\sum_{A=7}^{12} \mathcal{I}^{\alpha\beta}{}_{A} \mathcal{I}^{A}{}_{\mu\nu}
= 
\delta^{(\alpha}_{\mu} \delta^{\beta)}_{\nu} 
~&,&~~~
\sum_{A=13}^{17} \mathcal{I}^{\alpha\beta}{}_{A} \mathcal{I}^{A}{}_{\mu\nu}
=
\delta^{(\alpha}_{\mu} \delta^{\beta)}_{\nu}
-\frac{1}{3}\gamma^{\alpha\beta}\gamma^{\mu\nu}
~.
\ea
By means of this particular choice of intertwiner, the tensor fields $\varphi_1$ and $\varphi_2$ are symmetric, while $\varphi_3$ is symmetric and trace-free. We find that the parametrization 
\ba
(g_1(\varphi))^{\,\alpha\beta} &:=& \gamma^{\alpha\beta}+(\varphi_1)^{\alpha\beta}
,~~~ \label{Parametrization1}\\
(g_2(\varphi))^{\alpha\beta} &:=& \gamma^{\alpha\beta}
+ (\varphi_2)^{\alpha\beta}
,~~~\\
(g_3(\varphi))^{\,\alpha\beta} &:=&
(\varphi_3)^{\alpha\beta}
+ \sum_{n=1}^{\infty} (-1)^{n-1}\frac{1}{2^n}
\underbrace{\Big\{ \varphi_1, \Big\{ \varphi_1, \Big\{ \dots , \Big\{}_{\substack{n-1 \text{ anti-commutator}\\ \text{brackets}}}  \hspace{-0.1cm} \varphi_1 , 
\hspace{-0.2cm}
\underbrace{\big[ \varphi_1 , \varphi_3 \big]\vphantom{\Big\{ }}_{\substack{\text{commutator} \\ \text{bracket}}} \hspace{-0.2cm}\Big\} \dots \Big\} \Big\}\Big\}^{\alpha\beta} \qquad \label{Parametrization3}
\ea 
meets all symmetry and frame conditions, leaving us with an explicit expression of the canonical geometry in terms of the $17$ independent geometric degrees of freedom.

\section{Linearized gravitational dynamics}

The second step in the gravitational closure procedure is to set up and solve a set of countably many linear homogeneous partial differential equations---the {\it gravitational closure equations}---whose coefficients are determined by the principal tensor field expressed in terms of the canonical configuration variables. A derivation of these coefficients can be found in section IV.C of [\refcite{DSSW}]. Starting from general linear electrodynamics, the principal tensor field (\ref{principal_polynomial}), together with the parametrization (\ref{Parametrization1}-\ref{Parametrization3}), allows to determine these coefficients in terms of the canonical configuration variables $\{\varphi^1,\dots,\varphi^{17}\}$.
The gravitational closure equations then in turn determine the coefficient functionals $C_{A_1\dots A_N}[\varphi]$ of a power series ansatz for the gravitational Lagrangian density
\ba 
\mathcal{L}_{grav}[\varphi,k]=
\sum_{N=0}^{\infty} C_{A_1...A_N}[\varphi] \, k^{A_1}\dots k^{A_N}\,,
\ea 
where $k$ are the velocities associated with the canonical configuration variables $\varphi$.
Deriving gravitational field dynamics hence amounts to solving the gravitational closure equations---which turns out to be a difficult task for a non-metric geometry. However, we here only wish to study small area metric perturbations, which require just {\it linearized} gravitational field equations. 
In order to obtain the gravitational field equations underlying general linear electrodynamics to linear order in $\varphi$ and $k$, we only need to derive the Lagrangian density to second order
\ba
\mathcal{L}_{grav}[\varphi,k] = C[\varphi] + C_A[\varphi]k^A + C_{AB} [\varphi]k^A k^B + \mathcal{O}(3) \,.
\ea
That is, we only need to solve the gravitational closure equations for $C[\varphi]$ to second order, $C_A[\varphi]$ to first order and $C_{AB}[\varphi]$ to zeroth order in $\varphi$. We now also see why the parametrization (\ref{Parametrization1}-\ref{Parametrization3}) is particularly suited for this perturbative approach:
The configuration variables $\varphi$ can be directly employed as the perturbative degrees of freedom of an area metric perturbed around flat Minkowski spacetime $\eta=\text{diag}(1,-1,-1,-1)$.
We then perturbatively expand both the coefficients of the gravitational closure equations and the coefficient functionals $C_{A_1\dots A_N}[\varphi]$ in orders of $\varphi$, and evaluate the equations order by order. The such obtained gravitational Lagrangian, to second order in $\varphi$ and $k$, and the therefrom obtained linearized gravitational field equations can respectively be found in section III.D and IV.B of [\refcite{SSSW}]. These field equations can now be solved for specific matter distributions of interest.

\section{Solution around a point mass}

For instance, let us consider the Hamiltonian $H_{matter}$ of a point mass $M$ at rest, 
\ba
\frac{\delta H_{matter}}{\delta A(x)} = -M\delta^{(3)}(x)\,,
\ea
where $A$ denotes the perturbation of the lapse vector field.
Solving the linearized gravitational equations of motion for this particular small matter distribution yields the following area metric perturbations,
\ba
H^{0\beta 0\delta} &=& \gamma^{\beta\delta}(2A-\tfrac{1}{2} U -\tfrac{1}{2} V ) + \mathcal{O}(2) \label{H1}
\,,\\ 
H^{0\beta\gamma\delta} &=& \epsilon^{\beta\gamma\delta}(-A+\tfrac{3}{4} U + \tfrac{3}{4} V) + \mathcal{O}(2)  \label{H2}
\,,\\
H^{\alpha\beta\gamma\delta} &=& 2\gamma^{\alpha[\gamma}\gamma^{\delta]\beta}(U+2V) + \mathcal{O}(2) \label{H3}
\,,
\ea
where the scalar perturbations $A$, $U$ and $V$ can be expressed in terms of the Euclidean radial distance $r$ and undetermined integration constants $\alpha$, $\beta$, $\gamma$ and $\mu$:
\ba
U = \frac{M}{4\pi r}\left(\alpha + \beta e^{-\mu r}\right) , 
~~~
V = \frac{M}{4\pi r}\gamma e^{-\mu r} \,,
~~~
A = \frac{M}{16\pi r}\left(\alpha - \left(\beta + 3\gamma\right)e^{-\mu r}\right) . ~~~
\ea
Just like the gravitational and cosmological constants in the Einstein field equations of general relativity, these constants need to be determined by experiment.
For the principal tensor field of an area metric spacetime sourced by a point mass we obtain,
\ba 
P_G^{0000} = 1+2\Phi + \mathcal{O}(2) 
\,,
\quad 
P_G^{\alpha 000} = \mathcal{O}(2) 
\,,
\quad
P_G^{\alpha\beta 00} = -\tfrac{1}{6}\gamma^{\alpha\beta} + \mathcal{O}(2) 
\,,
\nonumber \\
P_G^{\alpha\beta\gamma 0} = \mathcal{O}(2)
\,,
\quad
P_G^{\alpha\beta\gamma\delta}  = (1-2\Phi)\gamma^{(\alpha\beta}\gamma^{\gamma\delta)} + \mathcal{O}(2) \,, \qquad\quad 
\label{point_mass_principal_polynomial}
\ea  
where we defined $\Phi := \frac{M}{4\pi r}\left[ -\frac{\alpha}{2} +\left(\frac{\beta}{2}+\frac{3\gamma}{4}\right) e^{-\mu r}\right]$.
In order to decide whether light rays split, one would need to determine the solution of the gravitational field equations for the area metric to second order\cite{GST}.
However, already to linear order, we find corrections to the usual linearized Schwarzschild metric in the form of a Yukawa potential. Allowing for vacuum birefringence, by assuming the electromagnetic field to be described by general linear electrodynamics, hence leaves its imprints on the spacetime geometry around a point mass. With the solution (\ref{point_mass_principal_polynomial}) one could now for instance study weak gravitational lensing in area metric spacetimes.

\section{Conclusions}

The weak gravitational field equations for the area metric spacetime geometry underlying general linear electrodynamics can be obtained by means of the gravitational closure mechanism. We find that modifying the spacetime geometry from a metric to an area metric, thereby allowing for vacuum birefringence, affects the canonical dynamical field equations and their solutions: For instance, for the weak field around a point mass in an area metric spacetime, we find first order Yukawa corrections to the Schwarzschild metric.
The linearized gravitational field equations now also enable further studies, e.g. the propagation of gravitational waves \cite{SSSW} or modifications of the standard Etherington relation\cite{SW} in area metric spacetimes.

\section*{Acknowledgments}
This paper is based on the research article [\refcite{SSSW}] and a talk presented in the Constructive Gravity session AT5 at the 15th Marcel Grossmann meeting. 
The author would like to thank F. P. Schuller for most valuable discussions and advice.
The author gratefully acknowledges support from the Studienstiftung des deutschen Volkes and the Ontario Trillium Scholarship (OTS) program.


\begin{thebibliography}{0}

\bibitem{HO} F. W. Hehl and Y. N. Obukhov, {\em Foundations of classical electrodynamics} (Birkh\"auser, 2003)

\bibitem{Rubilar1} G.F. Rubilar, 
{\em Annals\ Phys.} {\bf 11}, 717 (2002)

\bibitem{Rubilar2} G.F. Rubilar, Y.N. Obukhov, F. W. Hehl, 
{\em Int. J. Mod. Phys.} {D\,\bf 11}, 1227 (2002)

\bibitem{DSSW} M. D\"ull, F. P. Schuller, N. Stritzelberger and F. Wolz, 
{\em Phys.\ Rev.} {D\,\bf 97}, 084036 (2018)

\bibitem{GSWW} K. Giesel, F. P. Schuller, C. Witte and M. N. R. Wohlfarth, 
{\em Phys.\ Rev.} {D\,\bf 85}, 104042 (2012)

\bibitem{K} K. Kuchar, {\em J.\ Math.\ Phys.} {\bf 15}, 708 (1974)

\bibitem{HKT} S. A. Hojman, K. Kuchar and C. Teitelboim, 
{\em Annals\ Phys.} {\bf 96}, 88 (1976)

\bibitem{SSSW} J. Schneider, F. P. Schuller, N. Stritzelberger and F. Wolz, 
{\tt arXiv:hep-th/1708.03870} (2017)

\bibitem{KM} V. A. Kostelecky and M. Mewes, {\em Phys.\ Rev.} {D\,\bf 80}, 015020 (2009), 0905.0031

\bibitem{SWW} F. P. Schuller, C. Witte and M. N. R. Wohlfarth, 
{\em Annals\ Phys.} {\bf 325}, 9 (2010)

\bibitem{RRS} R. R\"atzel, S. Rivera and F. P. Schuller, 
{\em Phys.\ Rev.} {D\,\bf 83}, 044047 (2011)

\bibitem{GST} S. Grosse-Holz, F. P. Schuller and R. Tanzi, 
{\tt arXiv:hep-ph/1703.07183v2} (2017)

\bibitem{SW} F. P. Schuller and M. C. Werner, 
{\em Universe} {\bf 3}, 52 (2017)


\end{thebibliography}
\end{document}